\documentclass[letterpaper,twocolumn,prb,aps,superscriptaddress,floatfix,amsmath,amssymb]{revtex4}
\usepackage{graphicx}
\usepackage{multirow}
\usepackage{natbib}
\usepackage{array}
\usepackage{multirow,amssymb,amsbsy,amsmath,epstopdf}
\usepackage{color}
\usepackage{subfigure}
\usepackage{float}
\usepackage{upgreek}

\usepackage[colorlinks,citecolor=blue]{hyperref}

\begin{document}

\title{Temperature dependence of divacancy spin coherence in implanted silicon carbide}


\author{Wu-Xi Lin}
\affiliation{CAS Key Laboratory of Quantum Information, University of Science and Technology of China, Hefei 230026, People's Republic of China}
\affiliation{CAS Center for Excellence in Quantum Information and Quantum Physics, University of Science and Technology of China, Hefei 230026, People's Republic of China}

\author{Fei-Fei Yan}
\affiliation{CAS Key Laboratory of Quantum Information, University of Science and Technology of China, Hefei 230026, People's Republic of China}
\affiliation{CAS Center for Excellence in Quantum Information and Quantum Physics, University of Science and Technology of China, Hefei 230026, People's Republic of China}

\author{Qiang Li}\email{qianglee@ustc.edu.cn}
\affiliation{CAS Key Laboratory of Quantum Information, University of Science and Technology of China, Hefei 230026, People's Republic of China}
\affiliation{CAS Center for Excellence in Quantum Information and Quantum Physics, University of Science and Technology of China, Hefei 230026, People's Republic of China}

\author{Jun-feng Wang}
\affiliation{CAS Key Laboratory of Quantum Information, University of Science and Technology of China, Hefei 230026, People's Republic of China}
\affiliation{CAS Center for Excellence in Quantum Information and Quantum Physics, University of Science and Technology of China, Hefei 230026, People's Republic of China}

\author{Zhi-He Hao}
\affiliation{CAS Key Laboratory of Quantum Information, University of Science and Technology of China, Hefei 230026, People's Republic of China}
\affiliation{CAS Center for Excellence in Quantum Information and Quantum Physics, University of Science and Technology of China, Hefei 230026, People's Republic of China}

\author{Ji-Yang Zhou}
\affiliation{CAS Key Laboratory of Quantum Information, University of Science and Technology of China, Hefei 230026, People's Republic of China}
\affiliation{CAS Center for Excellence in Quantum Information and Quantum Physics, University of Science and Technology of China, Hefei 230026, People's Republic of China}

\author{Hao Li}
\author{Li-Xing You}
\affiliation{State Key Laboratory of Functional Materials for Informatics, Shanghai Institute of Microsystem and Information Technology, Chinese Academy of Sciences(CAS), Shanghai 200050, People's Republic of China}

\author{Jin-Shi Xu}\email{jsxu@ustc.edu.cn}
\affiliation{CAS Key Laboratory of Quantum Information, University of Science and Technology of China, Hefei 230026, People's Republic of China}
\affiliation{CAS Center for Excellence in Quantum Information and Quantum Physics, University of Science and Technology of China, Hefei 230026, People's Republic of China}

\author{Chuan-Feng Li}\email{cfli@ustc.edu.cn}
\affiliation{CAS Key Laboratory of Quantum Information, University of Science and Technology of China, Hefei 230026, People's Republic of China}
\affiliation{CAS Center for Excellence in Quantum Information and Quantum Physics, University of Science and Technology of China, Hefei 230026, People's Republic of China}

\author{Guang-Can Guo}
\affiliation{CAS Key Laboratory of Quantum Information, University of Science and Technology of China, Hefei 230026, People's Republic of China}
\affiliation{CAS Center for Excellence in Quantum Information and Quantum Physics, University of Science and Technology of China, Hefei 230026, People's Republic of China}
\date{\today}

\begin{abstract}
Spin defects in silicon carbide (SiC) have attracted increasing interest due to their excellent optical and spin properties, which are useful in quantum information processing. In this paper, we systematically investigate the temperature dependence of the spin properties of divacancy defects in implanted 4\emph{H}-SiC. The zero-field splitting parameter $D$, the inhomogeneous dephasing time $T_2^{*}$, the coherence time $T_2$, and the depolarization time $T_1$ are extensively explored in a temperature range from 5 to 300 K. Two samples implanted with different nitrogen molecule ion fluences ($\rm {N_2}^{+}$, $1\times 10^{14}/\rm cm^{2}$ and $1\times 10^{13}/\rm cm^{2}$) are investigated, whose spin properties are shown to have similar temperature-dependent behaviors. Still, the sample implanted with a lower ion fluence has longer $T_{2}$ and $T_{1}$. We provide possible theoretical explanations for the observed temperature-dependent dynamics. Our work promotes the understanding of the temperature dependence of spin properties in solid-state systems, which can be helpful for constructing wide temperature-range thermometers based on the mature semiconductor material.
\end{abstract}

\maketitle

\section{introduction}
Spin defects in solid-state materials have become an important candidate for qubits during the last two decades for their possible applications in quantum communication, quantum computation and quantum sensing \citep{AwschalomSum,AwschalomCompute,WrachtrupDiamond,AwschalomRoom,WrachtrupSingle,Dimond2,Dimond3,AwschalomPolytype,Astakhov,Castelletto,SiCAdd1,SiCAdd2,Lienhard,SiCAdd3,SiCAddLab}. Nitrogen-vacancy (NV) centers in diamond have been most frequently explored \citep{WrachtrupDiamond,Dimond2,Dimond3}. However, because the fabrication of suitable diamond-based devices for large-scale practical applications is still met with great challenges, much effort has been put into the search for similar spin defects in more mature solid-state materials. Silicon carbide is an outstanding semiconductor material that has been widely used in many electronic devices \citep{Material1,Material2} due to its excellent properties such as high thermal conductivity, chemical stability, and controllable electroconductivity. There are sophisticated technologies for the large-scale crystal growth of SiC and mature processes for the industrial production of SiC-based devices. So, the defects in SiC, which have been explored as bright single-photon sources \citep{Astakhov,Castelletto,Lienhard} and potential qubits \citep{AwschalomRoom,WrachtrupSingle,AwschalomPolytype,SiCAdd1,SiCAdd2,SiCAdd3,SiCAddLab}, have attracted great interest.

SiC has many polytypes and one of the most popular polytypes is 4\emph{H}-SiC, which has a high crystal quality and a layered hexagonal crystal structure \citep{AwschalomRoom}. One of the most frequently studied defects in 4\emph{H}-SiC is the divacancy defect consisting of a silicon vacancy and an adjacent carbon vacancy, which has spin $S=1$ \citep{AwschalomRoom,4HSiC}. There are two kinds of divacancies in 4\emph{H}-SiC, the \emph{c}-axis ones with $C_{3v}$ symmetry and the \emph{basal}-type ones with $C_{1h}$ symmetry \citep{AwschalomPolytype}. For a \emph{c}-axis divacancy, we mark its axis of symmetry as the \emph{c} axis and the direction of its spin is parallel to the \emph{c} axis. Recently, divacancies inside stacking faults have been shown to have excellent optical and spin properties at room temperature. Especially, the \emph{c}-axis divacancy named PL6 is shown to have bright photoluminescence fluorescence and a high spin readout contrast \citep{Li Qiang's paper}, which are comparable to those of NV centers in diamond.

In this paper, we systematically investigate the temperature dependence of the spin properties of the PL6 defects from 5 to 300 K in 4\emph{H}-SiC samples implanted by nitrogen molecule ion ($\rm {N_2}^{+}$) beams with two different ion fluences (the two samples are marked as sample A and sample B, where sample A is implanted with an ion fluence of $1\times 10^{14}/\rm cm^{2}$, and sample B is implanted with an ion fluence of $1\times 10^{13}/\rm cm^{2}$). The temperature-dependent behaviors of four important spin properties, the zero-field splitting (ZFS) parameter $D$, the inhomogeneous dephasing time $T_2^{*}$, the coherence time $T_2$, and the depolarization time $T_1$, are measured. $D$ decreases monotonically as the temperature increases. The inhomogeneous dephasing time $T_2^{*}$ fluctuates in a small time region at a wide range of temperatures. Although the coherence time $T_2$ decreases as the temperature increases, it is revived in a temperature range between 200 and 300 K. The inverse of the depolarization time $1/T_{1}$ shows an almost linear temperature-dependent property under 200 K and shows an approximately polynomial temperature dependence above 250 K. The temperature-dependent spin dynamics are similar for samples A and B implanted with these two different fluences of nitrogen molecule ions, but at the same temperature, the low-dose implanted sample B with a smaller PL6 concentration has longer $T_2$ and $T_1$. We discuss possible theoretical explanations for the observed temperature-dependent dynamics. This paper provides a profound understanding of the temperature-related behaviors of the divacancies in 4\emph{H}-SiC, which can be used for making wide temperature-range thermometers and designing quantum devices based on silicon carbide.


\section{Experimental Setup and Results}

\begin{figure}[!htb]
	\begin{center}
		\includegraphics[width=0.98\columnwidth]{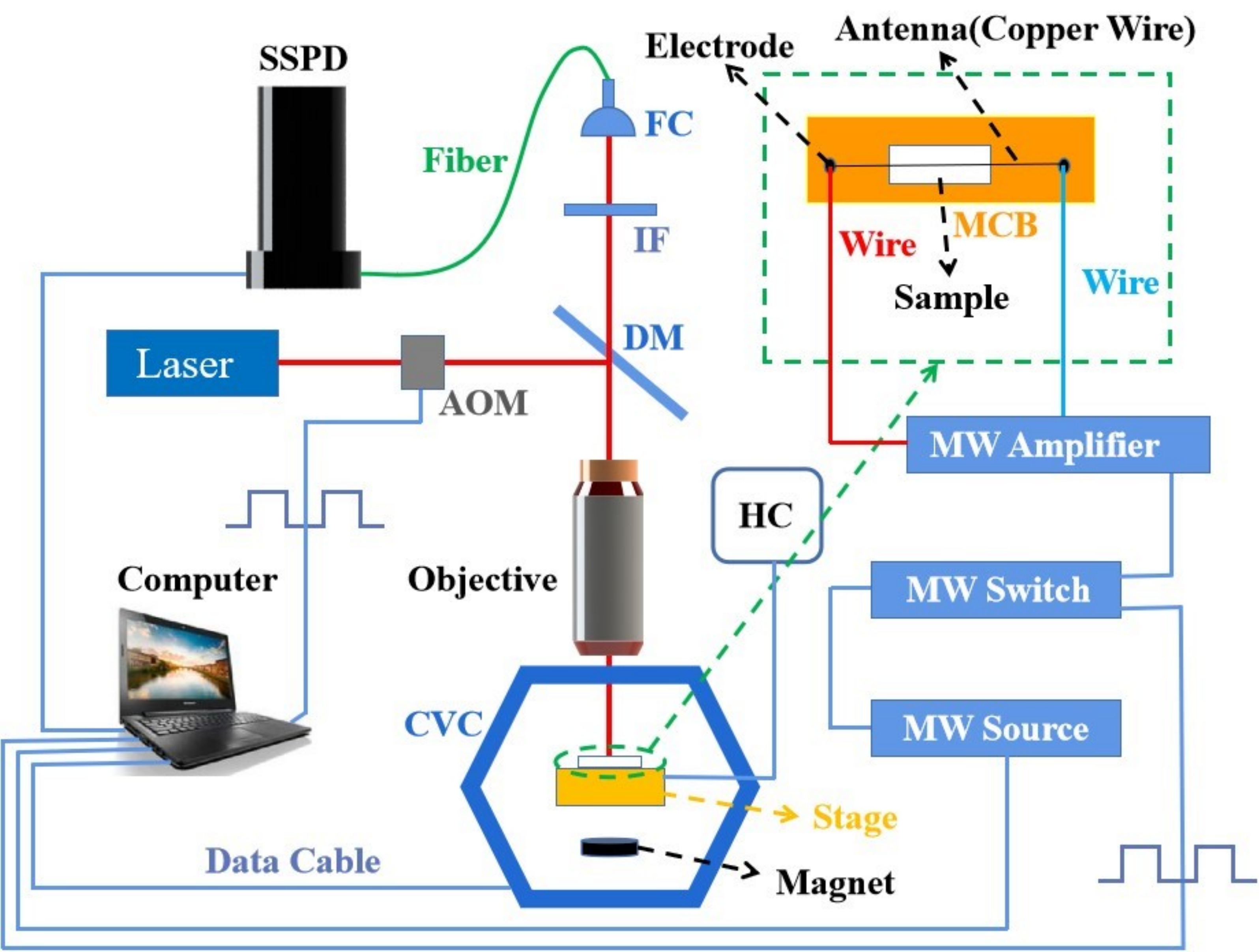}
		\caption{Experimental setup. The sample is mounted on a metal circuit board (MCB), which is attached to the stage in a cryogenic vacuum chamber (CVC). The temperature is controlled by a heater controller (HC). The pump laser modulated by an acousto-optic modulator (AOM) is focused by an objective to the sample. The fluorescence is collected by the same objective and filtered by a dichroic mirror (DM) and an interference filter (IF), which is then coupled by a fiber coupler (FC) and sent through the fiber to a superconducting single-photon detector (SSPD). The pump laser and microwave (MW) pulse sequences are controlled by a computer. The magnetic field is provided by a permanent magnet in the cryogenic chamber.}
		\label{fig1}
	\end{center}
\end{figure}

The experimental setup is shown in Fig.~\ref{fig1}. An infrared laser with a central wavelength of 920 nm modulated by an acousto-optic modulator (AOM) is used to pump the defect spins in the sample. The laser is focused on the sample by an near-infrared objective [Olympus, numerical aperture (NA) = 0.65] after being reflected by a 980-nm long-pass dichroic mirror (DM, Thorlabs). The fluorescence is collected by the same objective and further filtered by a 1000-nm long-pass interference filter (Thorlabs), then detected by a superconducting single-photon detector (Photon Technology). The sample is bonded on a metal circuit board with two electrodes, and we weld a 20-$\rm\upmu m$-diameter copper wire between the electrodes. The wire is attached to the surface of the sample. A microwave source (Mini-circuits, SSG6000) modulated by a microwave switch (Mini-circuits, ZASWA-2-50DR+) and amplified by a microwave amplifier (Mini-circuits, ZHL-30W-252+) is used to generate the microwave pulses that can manipulate the spin states of PL6 defects. The pulse sequences of the laser and microwave are controlled by a computer. The experimental results in this paper are obtained from the change of photoluminescence fluorescence excited by the laser between the two circumstances which are with or without the action of the microwave, which is marked as $\Delta \rm PL$ in arbitrary units (a.u.).

Samples are sliced from a commercially available high-purity 4\emph{H}-SiC epitaxial wafer with a 7-$\rm \upmu m$-thick epitaxial layer. According to some previous works, a hydrogen ion ($\rm H^{+}$) beam \citep{SiCAddLab} or carbon ion ($\rm C^{+}$) beam \citep{Li Qiang's paper} can be used to generate the PL6 defects in 4\emph{H}-SiC. However, nitrogen molecule ion ($\rm {N_2}^{+}$) implantation is rarely used to generate these kinds of defects in 4\emph{H}-SiC. But since $\rm {N_2}^{+}$ has a larger collision cross section than those of $\rm H^{+}$ and $\rm C^{+}$ when hitting the sample, we can reasonably guess that the $\rm {N_2}^{+}$ ion fluence can also effectively generate PL6. So, in this work, we use $\rm {N_2}^{+}$ ions with a kinetic energy of 30 keV to implant samples A and B with fluences of $1\times 10^{14}/\rm cm^{2}$ and $1\times 10^{13}/\rm cm^{2}$, respectively. After implantation, the samples are annealed at 1050$^\circ$C for 30 min. Under the above implantation and annealing conditions, the PL6 defects in 4\emph{H}-SiC can be efficiently generated. The PL6 concentration in sample B implanted with an $\rm {N_2}^{+}$ ion fluence of $1\times 10^{13}/\rm cm^{2}$ is smaller and this low-dose implanted sample is used for comparative experiments. 

The sample is mounted on the stage in a cryogenic vacuum chamber (Montana Instruments) with the temperature at around 4 K. The temperature is controlled by a heater whose temperature controlling range is between 5 and 300 K with an accuracy of 0.001 K. In order to split the approximately degenerate spin states, we put the sample in a static magnetic field, which is provided by a permanent magnet placed in this chamber. In the experiment, the magnetic field is set to be $1.80\times 10^{-2}$ T.

We first investigate the spin properties of sample A implanted by nitrogen molecule ions with a fluence of $1\times 10^{14}/\rm cm^{2}$. The ground-state spin Hamiltonian of PL6 defects can be written as \citep{WrachtrupSingle}
\begin{equation}
H=D[S_z^{2}-\frac{1}{3}S(S+1)]+E(S_x^{2}-S_y^{2})+g\mu_B\vec{S}\cdot \vec{B},
\end{equation}
where $g\approx 2$ is the electron's Lande-$g$ factor and $\mu_B$ is the Bohr magneton. $D$ and $E$ are both parameters describing the zero-field splitting which is caused by the electric dipole interaction between electrons, and they are related to the axisymmetric and non-axisymmetric part of the electric dipole interaction, respectively. For PL6, $D \gg E$ \citep{AwschalomPolytype}, and we only focus on $D$. The $S_x$, $S_y$, and $S_z$ are three components of the spin ($S=1$). There are two possible transitions between states (denoted as the magnetic quantum numbers) $\mathinner{|m_s=0\rangle} \leftrightarrow \mathinner{|m_s=+1\rangle}$ and $\mathinner{|m_s=0\rangle}\leftrightarrow\mathinner{|m_s=-1\rangle}$, respectively. As the applied \emph{c}-axis static magnetic field $\vec{B}$ is parallel to $\vec{S}$, the corresponding energy splittings which can be derived from the Hamiltonian are $\omega_1=D-\sqrt{(g\mu_B B/\hbar)^{2}+E^{2}}$ and $\omega_2=D+\sqrt{(g\mu_B B/\hbar)^{2}+E^{2}}$, so the ZFS parameter $D$ can be obtained by $D=\frac{1}{2}(\omega_1+\omega_2)$.

\begin{figure}[htbp]
	\begin{center}

	\includegraphics[width=0.98\columnwidth]{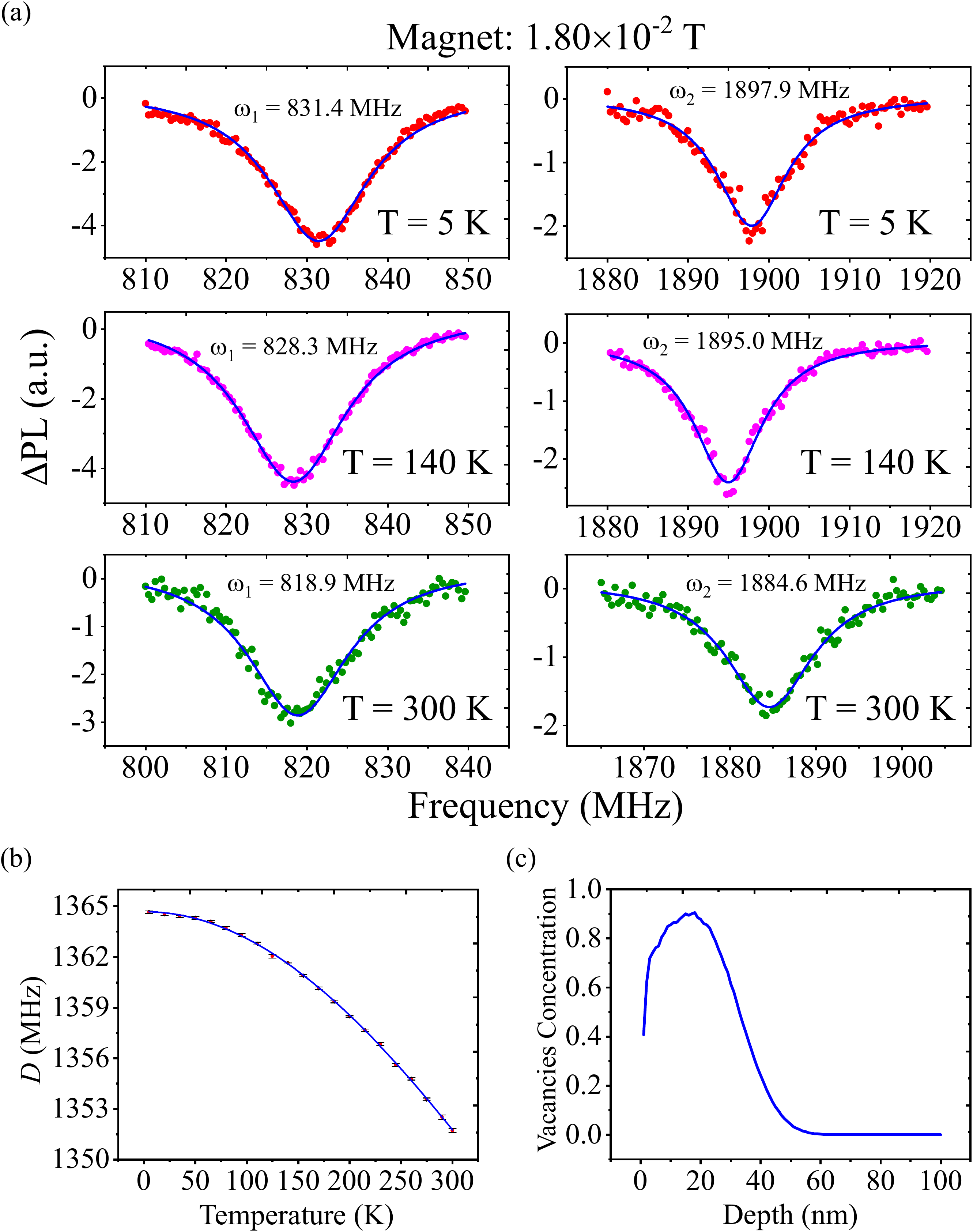}

	\caption{Temperature dependence of the ODMR spectrum and the ZFS parameter $D$ of sample A implanted with a $1\times 10^{14}/\rm cm^{2}$ nitrogen molecule ion fluence along with the simulation result of the vacancy defect's depth distribution by using SRIM. In (a) and (b) the dots are experimental data and the solid lines are fitting lines, while in (c) the solid line is the simulation curve. (a) The ODMR spectra of PL6 at 5, 140, and 300 K in a $1.80\times 10^{-2}$ T static magnetic field. The fitting lines are Lorentzian lines. $\omega_1$ and $\omega_2$ are the two corresponding resonant frequencies, respectively, and $D=(\omega_1+\omega_2)/2$. (b) Temperature dependence of $D$. The fitting line is the Debye-model fitting line with a determination coefficient of $0.9997$. $D$ decreases monotonically as the temperature increases. The error bars are deduced from the fitting errors. (c) The simulation result that shows the normalized vacancy concentration generated by the $\rm {N_2}^{+}$ ion implantation as a function of the depth from the sample surface. This result demonstrates that most vacancy defects, including PL6 defects, are distributed within 50 nm from the sample surface.}
	\label{fig2}
	\end{center}

\end{figure}

Employing optically detected magnetic resonance (ODMR) technology \citep{ODMR}, the resonant frequencies can be obtained. Figure~\ref{fig2}(a) shows three representative ODMR spectra at three different temperatures, 5, 140, and 300 K. The values of $\omega_1$ and $\omega_2$ are obtained from the Lorentzian fitting lines in the left and right panels, respectively. We will focus on the resonant frequency $\omega_1$ in all the coherent control experiments described below. Figure~\ref{fig2}(b) shows the relationship between the ZFS parameter $D$ and the temperature. The change of $D$ is about $-13$ MHz from 5 to 300 K. There are several different formulas that can be used to fit the curve shown in Fig.~\ref{fig2}(b): the Debye-model formula $D=[1304.1+60.6\exp(-2.7\times 10^{-6}K^{-2}T^{2})]\ \rm MHz$, which is derived from the simple Debye model theoretically \citep{D1}; the Varshni-form formula $D=(1364.7-0.2 K^{-1}\frac{T^{2}}{1348.2K+T})\ \rm MHz$ \citep{D2}; and the polynomial-form formula $D=(1364.6+3.5\times10^{-3}K^{-1}T-1.8\times10^{-4} K^{-2}T^{2}-1.5\times10^{-7} K^{-3}T^{3}+1.6\times10^{-9} K^{-4}T^{4}-2.7\times10^{-12} K^{-5}T^{5})\ \rm MHz$ \citep{D3,Important2}. For all the above fitting functions, the determination coefficients satisfy $R^{2}>0.99$, so all the functions are suitable for fitting the temperature dependence of $D$. Because of the comparatively more solid theoretical foundation \citep{D1}, we prefer to use the Debye-model formula to fit the $D$-temperature curve, which is shown in Fig.~\ref{fig2}(b). The temperature dependence of $D$ can be used in the high-sensitivity quantum sensing of temperature \citep{Important2,D2,D3,Thermal1,Thermal2}.

Because the temperature dependence of the spin properties of the defects can be affected by their distribution, we also use the stopping and range of ions in matter (SRIM) \citep{SRIM_1,SRIM_2} to simulate the depth distribution of the vacancy defects (including PL6) generated by the 30-keV $\rm {N_2}^{+}$ implanted ions. Figure~\ref{fig2}(c) shows the simulation result of the normalized concentration of vacancy defects as a function of the distance to the sample surface. Deduced from the result, most of the vacancy defects, including the PL6 defects, are distributed within 50 nm from the surface of the sample.



\begin{figure}[!htb]
	\begin{center}
		\includegraphics[width=0.98\columnwidth]{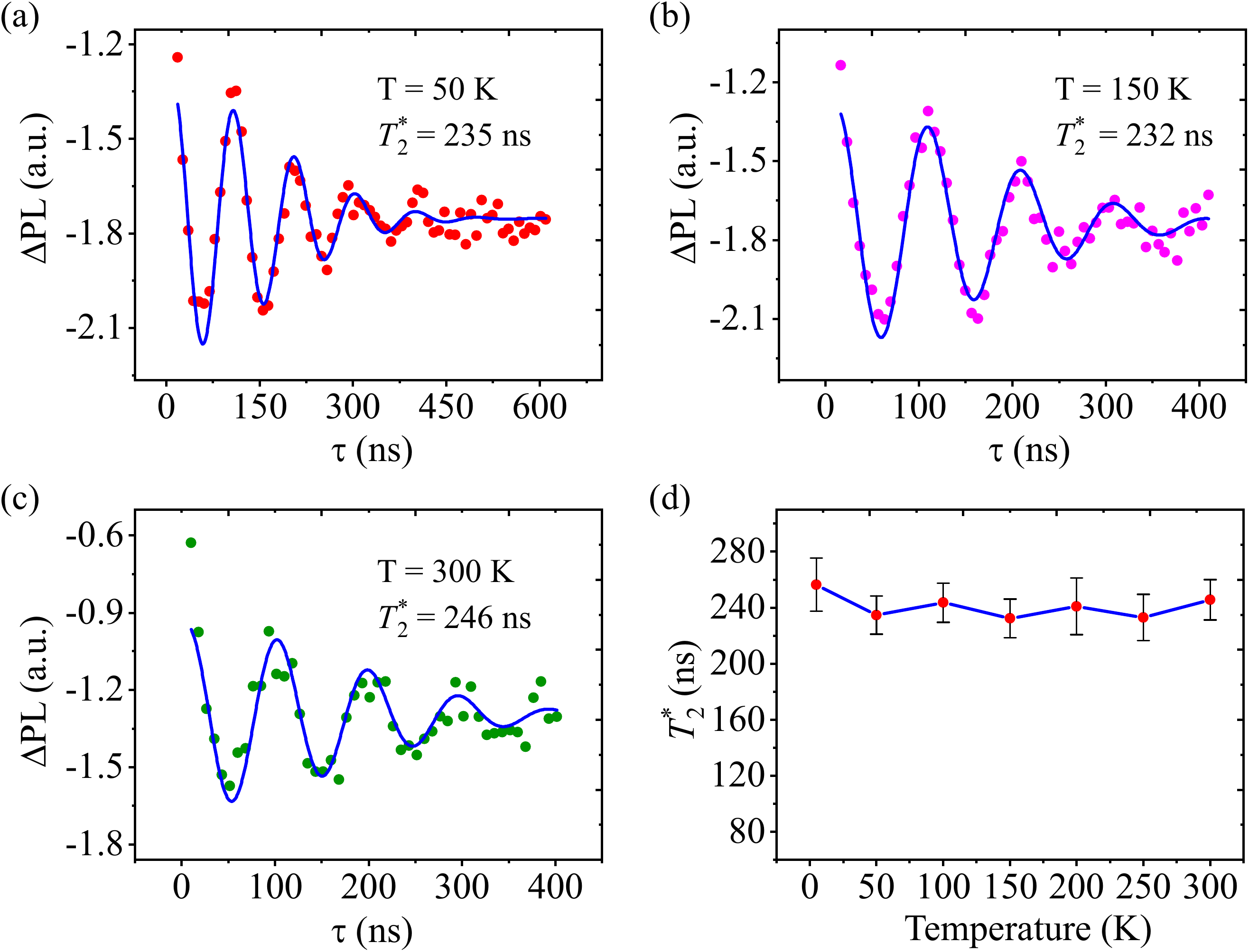}
		\caption{Temperature dependence of Ramsey interference and the inhomogeneous dephasing time $T_2^{*}$ of sample A implanted with a $1\times 10^{14}/\rm cm^{2}$ nitrogen molecule ion fluence. (a)-(c) Ramsey interference curves at 50, 150, and 300 K with $T_2^{*}$ derived from the fitting formula, respectively. The dots are experimental data and the solid lines are the fitting lines. (d) Temperature dependence of $T_2^{*}$ from 5 to 300 K. $T_2^{*}$ remains stable at different temperatures. Error bars are deduced from the fitting errors.}
		\label{fig3}
	\end{center}
\end{figure}

We then measure the inhomogeneous dephasing time $T_2^{*}$ of PL6 at different temperatures inferred from the Ramsey interference curve. $T_2^{*}$ is an important factor that affects the performance of quantum sensors based on Ramsey interference \citep{Important2,Thermal1}. The microwave pulse sequence used in the Ramsey interference experiments is $\pi/2-\tau-\pi/2$. $\pi/2$ means the duration time of the microwave is $\pi/(2\Omega_R)$, where $\Omega_R$ is the angular frequency of the Rabi oscillation \citep{Rabi} of the spin system under the action of the microwave with the same power. $\tau$ is the length of the time interval during which the system is allowed to evolve freely without the action of the microwave. We set the detuning frequency of the microwave as $\delta=10\ \rm MHz$. Figure~\ref{fig3}(a)-\ref{fig3}(c) show three representative Ramsey interference curves at 50, 150, and 300 K, respectively. We fit these curves by the formula $a\exp[-(\tau/T_2^{*})^{2}]\cos(2\pi \delta \tau+\phi)+b$, where $a$, $b$, and $\phi$ are free parameters, and we take the exponent of $\tau/T_2^{*}$ as 2 because of a reasonable hypothesis that the noise which causes the signal to decay has a Gaussian distribution, as shown in the former theoretical results \citep{Ramsey,Ramsey2}. $T_2^{*}$ can then be derived from the fitting formula. Figure~\ref{fig3}(d) shows the temperature dependence of $T_2^{*}$, from which we can find that $T_2^{*}$ remains almost the same. The collapse of the Ramsey interference oscillation is mainly due to the interaction between the defect spins and the nuclear spins, the fluctuation of the external magnetic field, and the instability of the frequency of the microwave. The increase in temperature mainly influences the intensity of the lattice vibration, which does not much affect the behavior of inhomogeneous dephasing.

\begin{figure}[!htb]
	\begin{center}
		\includegraphics[width=0.98\columnwidth]{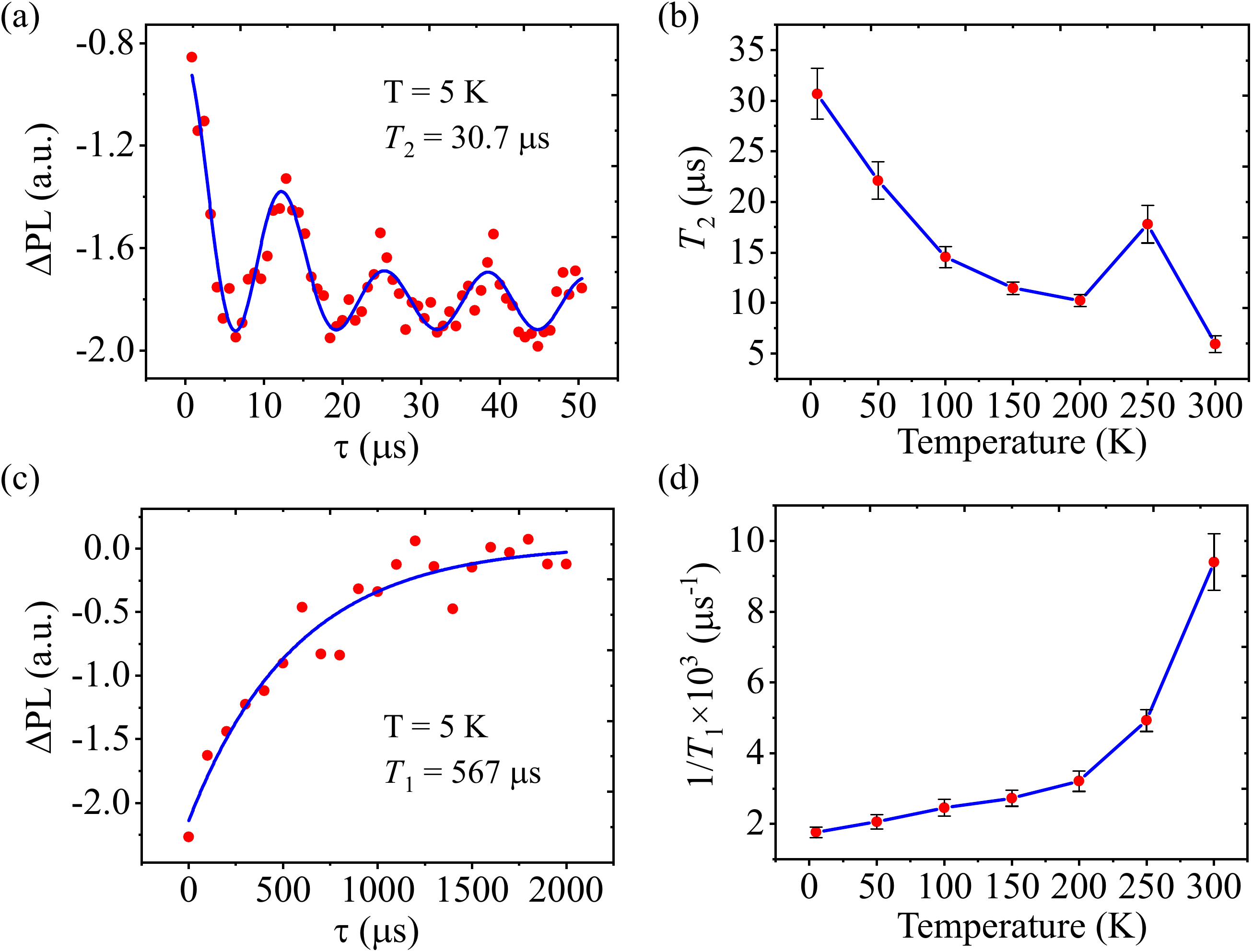}
		\caption{Temperature dependence of the coherence time $T_2$ and the depolarization time $T_1$ of sample A implanted with a $1\times 10^{14}/\rm cm^{2}$ nitrogen molecule ion fluence. (a) The spin echo oscillation at 5 K with $T_2=30.7\ \rm\upmu s$ derived from the fitting formula (the solid line).  (b) Temperature dependence of $T_2$ from 5 to 300 K. An anomalous revival of $T_2$ occurs at around 250 K. (c) The depolarization curve at 5 K with $T_1=567\ \rm\upmu s$ derived from the fitting formula (the solid line). (d) Temperature dependence of $1/T_1$ from 5 to 300 K, which transits from a linear relationship to a polynomial relationship. The dots are experimental data. Error bars are deduced from the fitting errors.}
		\label{fig4}
	\end{center}
\end{figure}

Since the coherence time $T_2$ is critical in the spin-based quantum technology \citep{Divincenzo}, we further measure its temperature-dependent behavior. The standard method of measuring $T_2$ is to measure the spin echo oscillation, and the microwave pulse sequence used in this experiment is $\pi/2-\tau/2-\pi-\tau/2-\pi/2$. Figure~\ref{fig4}(a) shows a representative spin echo curve at 5 K. Because of the existence of the static magnetic field, the spin echo attenuated oscillation is an electron spin echo envelope modulation (ESEEM) curve \citep{Nuclear1,Nuclear2,Nuclear3}. The experimental data are fitted by the formula $a\exp(-\tau/T_2)[1-b\sin^{2}(\pi f_1\tau)][1-c\sin^{2}(\pi f_2\tau)]+d$, where $a$, $b$, $c$, and $d$ are free parameters. $f_1$ and $f_2$ are Larmor precession frequencies of the spin magnetic moments of the $^{13}$C and $^{29}$Si nuclei in the static magnetic field. Figure~\ref{fig4}(b) shows the temperature dependence of $T_{2}$. We can find an unusual phenomenon, that is, in a specific temperature range between 200 and 300 K, $T_2$ becomes larger as the temperature increases. But usually the coherence time of the defects in solid-state systems decreases monotonically as the temperature increases. Similar behavior was observed from silicon vacancy spins in 4\emph{H}-SiC \citep{Important1}.

In order to explain this unusual nonmonotonic relationship between $T_2$ and temperature, we assume that it may be attributed to the dynamic Jahn-Teller effect \citep{Important1}. In the 4\emph{H}-SiC crystal, there are carbon and silicon atoms adjacent to the divacancy. At low temperatures, orbital pairing occurs between neighboring atoms' electrons and leads to the Jahn-Teller distortion, which defines the structure of the divacancy. As the temperature increases to a specific value, the energy of the thermal vibration is large enough for unraveling the orbital pairs, so rapid thermally activated reorientation occurs and the Jahn-Teller distortion disappears. The restructuring of the divacancy induced by the orbital pairs' unraveling can possibly lead the coherence time to increase and we can find the abnormal phenomenon that $T_2$ is longer at a higher temperature than at a lower temperature.


The coherence time $T_2$ of defects is restricted by the depolarization time $T_1$, which is given by $T_2<0.5T_1$ \citep{T2T11} or $T_2<1T_1$ \citep{T2T12}. We measure the temperature-dependent behavior of $T_1$. The spin state is first polarized and then evolves freely for a period of time $\tau$, from which the photoluminescence fluorescence is detected. A similar detection is then performed in the case with a $\pi$ microwave pulse after the polarizing process. The fluorescence difference between these two processes is measured as the function of $\tau$. Figure~\ref{fig4}(c) shows a representative depolarization curve at 5 K. The experimental data are fitted by the exponential decay formula $a\exp(-\tau/T_1)+b$, where $a$ and $b$ are free parameters. $T_1$ can then be derived from the fitting formula. Figure~\ref{fig4}(d) shows the temperature dependence of $1/T_1$. In the temperature region from 5 to 200 K, $1/T_1$ increases almost linearly as the temperature increases. Two main factors are involved to cause the spin depolarization, including the direct energy transition between the defect spins and the phonons, and the Orbach mechanism. $1/T_1$ can be described by $1/T_1\approx a\coth(E_1/k_BT)+b/[\exp(E_2/k_BT)-1]+c$ \citep{T1theo1,T1theo2}, where $k_B$ is the Boltzmann's constant and $c$ is a parameter independent of temperature. $E_1$ is proportional to the energy splitting between the spin states, and it satisfies $k_BT \gg E_1$ above 1 K. $E_2$ is proportional to the energy splitting between the orbital ground state and a nearby orbital excited state, which always satisfies $k_BT \ll E_2$ below 300 K. So we have $\coth(E_1/k_BT)\approx k_BT/E_1$ and $b/[\exp(E_2/k_BT)-1]\approx 0$ and the relationship becomes $1/T_1\approx a'T+c'$. When the temperature is above 250 K, the depolarization of the spin state is mainly caused by the two-phonon Raman process, which can be described by \citep{T1theo1,T1theo2} $1/T_1\approx aT^s+bT^{s+1}+cT^{s+2}$. $s=2d-1$ and $d$ is the dimension of the sample with the main term being $aT^s$. Because PL6 defects are mainly distributed near the sample surface [as the simulation result displayed in Fig.~\ref{fig2}(c) shows], we can approximately take $d=2$ and have $1/T_1\propto T^3$. From the experimental result we find that $[T_1(T=250K)/T_1(T=300K)]^{1/3}\approx 1.9^{1/3}\approx1.24\approx300K/250K$, which is approximately consistent with the theoretical analysis. The linear-form temperature dependence of $1/T_{1}$ transits to the polynomial-form temperature dependence in the temperature range between 200 and 250 K. For this sample A implanted with the $1\times 10^{14}/\rm cm^{2}$ ion fluence, the longest $T_{1}$ is measured to be about 500 $\rm\upmu s$, while the longest $T_{2}$ is about 30 $\rm\upmu s$. The coherence time can be increased further by using the dynamical decoupling method \citep{Science 330} and increasing the quality of the crystals, for example, reducing the defects in the samples and using the isotopic purification samples \citep{NatureMat 19, NatureCom 11}.

\begin{figure}[!htb]
	\begin{center}
		\includegraphics[width=0.98\columnwidth]{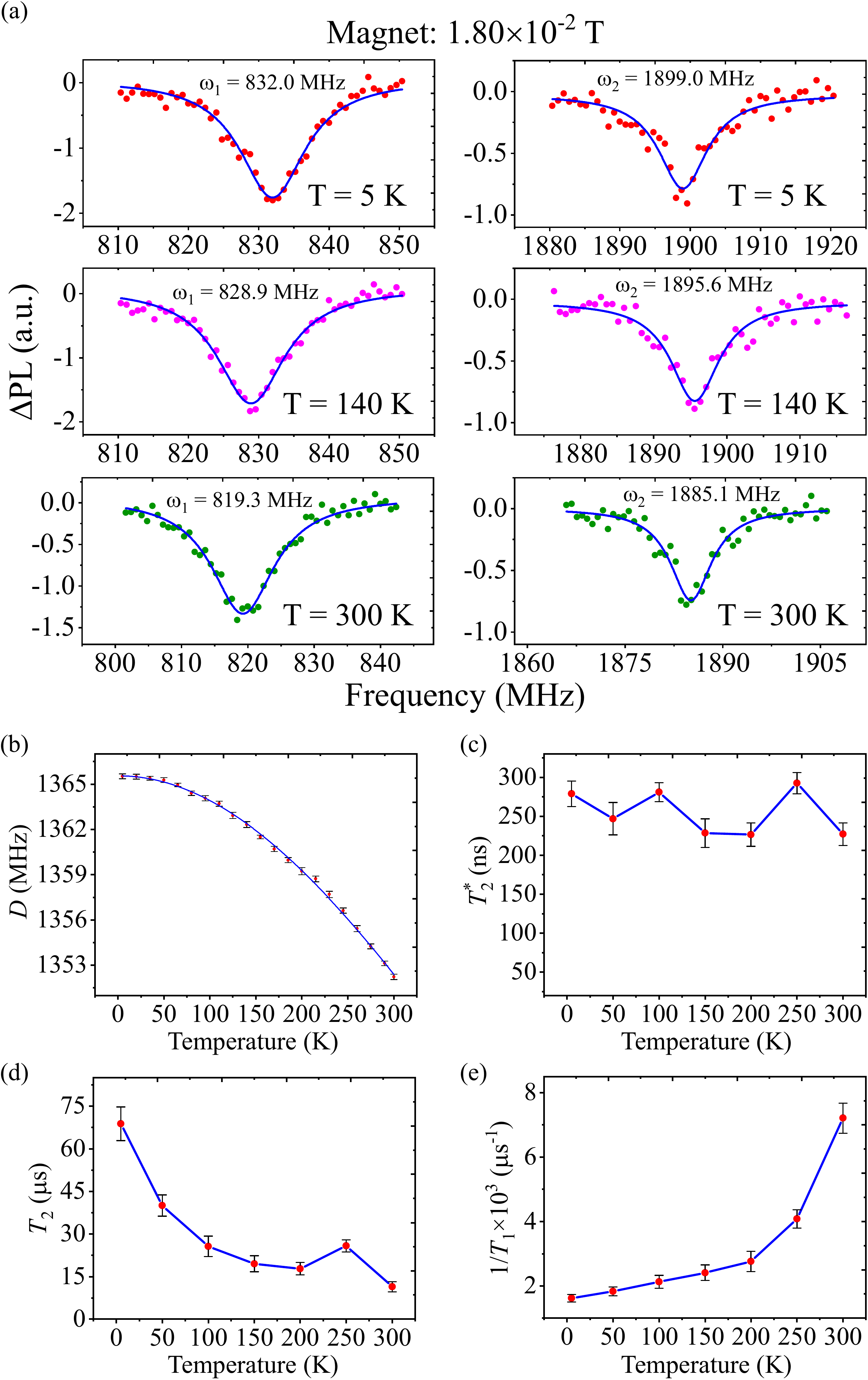}
		\caption{Experimental results of sample B implanted with a nitrogen molecule ion fluence of $1\times 10^{13}/\rm cm^{2}$. (a) Typical ODMR spectra at 5, 140, and 300 K. The ODMR signals of the low-dose implanted sample B have smaller intensities, smaller signal-to-noise ratios, and narrower full width at half maxima (FWHM) than those of the high-dose implanted sample A. (b) Temperature dependence of $D$. The solid line is the Debye-model fitting curve with the determination coefficient $R^{2}=0.9986$. (c) Temperature dependence of $T_2^{*}$. $T_2^{*}$ remains stable at different temperatures. (d) Temperature dependence of $T_2$. An anomalous revival of $T_2$ occurs at around 250 K. (e) Temperature dependence of $1/T_1$, which transits from a linear relationship to a polynomial relationship. At the same temperature, sample B has longer $T_2$ and $T_1$ than those of sample A implanted with an ion fluence of $1\times 10^{14}/\rm cm^{2}$. Error bars are deduced from the fitting errors.}
		\label{fig5}
	\end{center}
\end{figure}

Finally, we provide the experimental results of PL6 defects in sample B implanted with a $1\times 10^{13}/\rm cm^2$ nitrogen molecule ion fluence. Figure~\ref{fig5}(a) shows typical ODMR spectra of this low-dose implanted sample B at 5, 140, and 300 K. The ODMR spectra in Figs.~\ref{fig2}(a) and \ref{fig5}(a) are normalized by the same normalization factor. From the two sets of ODMR spectra we can find that the ODMR signal intensity of sample B is about one-third to one-half of the ODMR signal intensity of sample A at the same temperature, which shows that a lower dose of implanted ions can indeed produce a smaller concentration of PL6 defects. We can also find that the low-dose implanted sample B's ODMR signal has a lower signal-to-noise ratio and a narrower full width at half maximum (FWHM) than those of the high-dose implanted sample A's ODMR signal at the same temperature. Figure~\ref{fig5}(b) shows the temperature dependence of the ZFS parameter $D$, which can also be well fitted by the Debye-model formula $D=[1301.3+64.3\exp(-2.5\times10^{-6}K^{-2}T^{2})]\ \rm MHz$ \citep{D1}, Varshni-form formula $D=(1365.6-0.2K^{-1}\frac{T^{2}}{1268.3K+T})\ \rm MHz$ \citep{D2}, and polynomial-form formula $D=(1365.4+1.2\times10^{-2}K^{-1}T-3.5\times10^{-4}K^{-2}T^{2}+8.8\times10^{-7}K^{-3}T^{3}-4.2\times10^{-10}K^{-4}T^{4}-2.2\times10^{-12}K^{-5}T^{5})\ \rm MHz$ \citep{D3,Important2}, respectively. The fitting curve that we show in Fig.~\ref{fig5}(b) is the Debye-model curve. $D$ also changes about $-13$ MHz from 5 to 300 K. Figure~\ref{fig5}(c) shows the temperature dependence of $T_2^{*}$, from which we can find that $T_2^{*}$ still remains stable for this low-dose implanted sample B. The value of $T_2^{*}$ is almost not affected by the implanted ion fluence. An abnormal nonmonotonic behavior is still observed in the temperature dependence of $T_2$ in the temperature range between 200 and 300 K, which is shown in Fig.~\ref{fig5}(d). The values of $T_2$ are almost two times those shown in Fig.~\ref{fig4}(b), which may due to the smaller total dipole interaction between the defect spins and environmental nuclear spins in this low-dose implanted sample B with a smaller defect concentration. Figure~\ref{fig5}(e) shows the temperature dependence of $T_1$. The behavior is similar to that shown in Fig.~\ref{fig4}(d), but this low-dose implanted sample B has a longer depolarization time. The sample with a higher defect concentration exists with larger resonant magnetic or electric noises, which depolarize the spin states \citep{T1concen1,T1concen2,T1concen3}.

\section{Conclusion}
In this paper, we explore the temperature dependence of PL6 defect spin properties in 4\emph{H}-SiC implanted with two different nitrogen molecule ion fluences between 5 and 300 K. The ZFS parameter is shown to decrease monotonically as the temperature increases. It changes about $-13$ MHz from 5 to 300 K, which can be used for constructing wide temperature-range thermometers. The inhomogeneous dephasing time is shown to be stable and remains almost the same in the two samples implanted with different ion fluences. Although the evolution trends are similar for both samples, the coherence time and depolarization time are longer in the sample implanted with a lower ion fluence. An increase in the coherence time is observed during the increase of temperature in both samples, which is attributed to the disappearance of the Jahn-Teller distortion. We discuss possible theoretical explanations for the observed phenomena. Our work is useful in the understanding of temperature-related spin dynamics in SiC, and will stimulate further investigations in this area.


\section*{ACKNOWLEDGMENTS}
This work was supported by the National Key Research and Development Program of China (Grants No. 2016YFA0302700), the National Natural Science Foundation of China (Grants No.\ U19A2075, 61725504, 61905233, 11774335, 11821404 and 11975221), the Key Research Program of Frontier Sciences, Chinese Academy of Sciences (CAS) (Grant No. QYZDY-SSW-SLH003), the Anhui Initiative in Quantum Information Technologies (Grants No. AHY060300 and No. AHY020100), and the Fundamental Research Funds for the Central Universities (Grants No.\ WK2030380017 and WK2470000026), the National Postdoctoral Program for Innovative Talents (Grant No.\ BX20200326). This work was partially performed at the University of Science and Technology of China Center for Micro and Nanoscale Research and Fabrication.

W.-X. L. and F.-F. Y. contributed equally to this work.

\end{document}